\documentclass[
aps,
prd,
superscriptaddress,
amsfonts,
amssymb,
amsmath,
nopreprintnumbers,
floatfix,
nofootinbib,
unsortedaddress,
noshowpacs,
tightenlines,
floats,
letterpaper,
eqsecnum,
]{revtex4}

\begin{document}

\title[
Various Hamiltonian formulations of $ f(\mathcal R) $ gravity and their canonical relationships
]{
Various Hamiltonian formulations of $ f(\mathcal R) $ gravity and their canonical relationships
}

\author{Nathalie Deruelle}
\affiliation{
APC, UMR 7164 du CNRS, Universit\'e Paris 7, 
75205 Paris, France
}

\author{Yuuiti Sendouda}
\affiliation{
Yukawa Institute for Theoretical Physics, Kyoto University,
Kyoto 606--8502, Japan
}

\author{Ahmed Youssef\,}
\affiliation{
APC, UMR 7164 du CNRS, Universit\'e Paris 7, 
75205 Paris, France
}

\date{
\today
}

\begin{abstract}
Various Hamiltonian formulations of $f(\mathcal R)$ gravity can be found in the literature.
Some authors follow the Ostrogradsky treatment of higher derivative theories and introduce as extra variables {\it first} order time derivatives of the metric (typically the extrinsic curvature).
Some others take advantage of the conformal equivalence of $f(\mathcal R)$ theory with Einstein's gravity coupled to a scalar field and introduce as an extra variable the scalar curvature $\mathcal R$ itself, which includes {\it second} time derivatives of the metric.
We show that, contrarily to some claims, these formulations are related by canonical transformations.
\end{abstract}


\maketitle

\section{
Introduction
}

A currently fashionable class of extended theories of gravity are the so-called $f(\mathcal R)$ theories whose Lagrangian is an arbitrary function of the scalar curvature $\mathcal R$ rather than simply $\mathcal R$ as in General Relativity, see e.g.\ \cite{Sotiriou:2008rpLobo:2008sgSotiriou:2008veFaraoni:2008mf} for recent reviews.
``Metric'' $f(\mathcal R)$ gravity\footnote{See \cite{Sotiriou:2008rpLobo:2008sgSotiriou:2008veFaraoni:2008mf} and references therein for variations ``\`a la Palatini.''} has two remarkable features~:
it is a ``higher derivative'' theory, that is, the field equations are fourth order differential equations for the metric~;
and these field equations are conformally equivalent to Einstein's equations minimally coupled to a scalar field \cite{Teyssandier:1983zz}.
This means that it possesses one extra degree of freedom, beyond those of Einstein's gravity \cite{Stelle:1977ry}.

The first Hamiltonian formulation of $f(\mathcal R)$ gravity, more precisely of $\mathcal R^2$\,, was performed by Boulware \cite{Boulware:1983yj} who chose as the extra degree of freedom the scalar curvature itself, that is, a function of {\it second} time derivatives of the metric.
Many authors subsequently followed this route, see e.g.\ \cite{Hawking:1984ph,Schmidt:1994iz,Demaret:1995wp,Deruelle:prep,Sendouda:prep}.

In parallel, an alternative Hamiltonian formulation of $f(\mathcal R)$ gravity was initiated by Buchbinder and Lyahovich \cite{Buchbinder:1987vp}, based on the ``Ostrogradsky procedure'' (see e.g.\ \cite{Woodard:2006nt} for a vivid review), which consists in promoting to the status of independent variable {\it first} order time derivatives of the metric (typically the extrinsic curvature).
For developments along this line, see e.g.\ \cite{Querella:1998ke,Ezawa:1998ax,Ezawa:2003wh,Ezawa:2005zr,Ezawa:2009rh,Capozziello:2007gm}.

Schmidt \cite{Schmidt:1994iz} clearly differentiated these alternative formulations, which are sometimes put on the same footing, see e.g.\ \cite{Demaret:1995wp}.
Now, since they must both yield the same equations of motion, one expects that they should be equivalent, that is, related by a canonical transformation.
However Ezawa {\it et al.}, \cite{Ezawa:1998ax,Ezawa:2009rh}, claim that they are {\it not}~: we shall show in this paper that they are.

We thus generalize to $f(\mathcal R)$ gauge field theories the result obtained in \cite{Andrzejewski:2007by} in the simple case of $L(q,\dot q,\ddot q)$ Lagrangians.
Moreover, in giving the explicit, highly non-linear, form of the transformation, we make it clear that this equivalence may hold at the classical level only (a point already made by Schmidt \cite{Schmidt:1994iz} in the simple case of minisuperspace).
\vskip 0.3cm

The paper is organized as follows.

We start in Section~\ref{sec:E} by recalling the Arnowitt--Deser--Misner (ADM) Hamiltonian formulation of $f(\mathcal R)$ gravity in the Einstein frame, where it is equivalent to General Relativity  \cite{Arnowitt:1962hi}.
This will serve as our ``Rashid stone''\footnote{Also known as the ``Rosetta stone.''} to evaluate subsequent formulations and fix the notations.

In the original, Jordan frame, $f(\mathcal R)$ gravity is explicitly a higher derivative theory.
In Section~\ref{sec:J} we present a Hamiltonian formulation in Boulware's line, promoting $\mathcal R$ to the status of independent variable.
As far as we know this treatment is new, and extends those of \cite{Hawking:1984ph,Schmidt:1994iz,Demaret:1995wp}.

In Section~\ref{sec:O} we turn to a Hamiltonian formulation ``\`a la Ostrogradsky,'' taking the trace of the extrinsic curvature as the extra degree of freedom.
We believe our treatment is simpler than those presented in \cite{Querella:1998ke,Ezawa:1998ax,Ezawa:2003wh,Ezawa:2005zr,Ezawa:2009rh}.

Section~\ref{sec:trans} is the core of the paper, where we explicitly exhibit the canonical transformations which turn the Einstein frame Hamiltonian into the Jordan frame one and then into the Ostrogradsky one.

Section~\ref{sec:conc} summarizes our results.

A number of self-contained Appendices complement and illustrate the core of the paper.
In Appendix~\ref{sec:toy} we derive in detail the Hamiltonians associated to a toy higher derivative Lagrangian of the type $L=L(q,\dot q,\ddot q)$\,, when either $\dot q$ or $\ddot q$ is taken as a new variable.
Appendix~\ref{sec:J2E} is a short recap of the conformal equivalence of the Einstein versus Jordan frame formulations of $f(\mathcal R)$ gravity.
Finally Appendix~\ref{sec:mini} applies our general results to the simple case of minisuperspace.

\section{\label{sec:E}
Einstein frame Hamiltonian of $f(\mathcal R)$ gravity
}

This Section summarizes the ADM formalism \cite{Arnowitt:1962hi} and fixes some notations (see e.g.\ \cite{Misner:1974qy,Wald:1984rg,Gourgoulhon:2007ue,Sendouda:prep} for a more geometrical approach).
\\

Consider a four dimensional manifold $\mathcal M$ whose points are labelled by some arbitrary ``ADM'' coordinates $x^i$ with $i=\{0,1,2,3\}$\,, and endowed with a metric $\tilde g_{ij}(x^k)$ with signature $(-,+,+,+)$\,, determinant $\tilde g$ and associated covariant derivative $\tilde\nabla_i$\,.
Suppose that $\mathcal M$ can be foliated by a family of spacelike $ 3 $-surfaces $\Sigma_t$\,, defined by $t=x^0$\,.
Let $\tilde h_{ab}\equiv\tilde g_{ab}|_{x^0=t}$ with $a, b$ running from 1 to 3 be the metric on $\Sigma_t$\,, $\tilde h$ its determinant, $\tilde h^{ab}$ its inverse and denote by $\tilde D_a$ the associated covariant derivative.
Three basis vector fields on $\Sigma_t$ are $\partial_a$\,, with components $\delta^i_a$~;
introduce too the future-pointing unit normal vector $\tilde n$ to the surface $\Sigma_t$\,, that is, to the three vectors $\partial_a$~;
its components are $\tilde n_a=0$\,, $\tilde n_0=-1/\sqrt{-\tilde g^{00}}$~;
$\tilde n^0=\sqrt{-\tilde g^{00}}$\,, $\tilde n^a=-\tilde g^{0a}/\sqrt{-\tilde g^{00}}$\,.
Decompose then the time-like basis vector $\partial_0$ (with components $\delta^i_0$) on the normal vector and the three basis vectors $\partial_a$~:
$\delta^i_0=\tilde N\tilde n^i+\tilde N^a\delta^i_a$~;
$\tilde N=1/\sqrt{-\tilde g^{00}}$ and $\tilde N^a=-\tilde g^{0a}/\tilde g^{00}$ are the ``lapse'' and ``shift.''
Together with the induced metric $\tilde h_{ab}$ they constitute the ``ADM variables.''
In terms of these variables we have $\sqrt{-\tilde g}=\tilde N\sqrt{\tilde h}$\,, $\tilde n^0=1/\tilde N$\,, $\tilde n^a=-\tilde N^a/\tilde N$\,, and the components of the $ 4 $-metric read
\begin{equation}
\left\{
\begin{aligned}
 &
\tilde g_{00}
 = -\tilde N^2 + \tilde N_a\tilde N^a\,,
\quad
\tilde g_{0a}
 = \tilde N_a\,,
\quad
\tilde g_{ab}
 = \tilde h_{ab}\,, \\
 &
\tilde g^{00}
 = -\frac{1}{\tilde N^2}\,,
\quad
\tilde g^{0a}
 = \frac{\tilde N^a}{\tilde N^2}\,,
\quad
\tilde g^{ab}
 = \tilde h^{ab} - \frac{\tilde N^a\tilde N^b}{\tilde N^2}\,.
\end{aligned}
\right.
\label{eq:EADM}
\end{equation}
(Here and in the following indices of three dimensional objects are moved with the induced metric.)
Introduce finally the extrinsic curvature of $\Sigma_t$~:
\begin{equation}
\tilde K_{ab}
 \equiv
   \tilde\nabla_a \tilde n_b
 = \frac{1}{2\,\tilde N}\,
   (\dot{\tilde h}_{ab} - \tilde D_a \tilde N_b - \tilde D_b \tilde N_a)\,,
\label{eq:EdefK}
\end{equation}
where a dot denotes a time derivative~: $\dot{\tilde h}_{ab}=\frac{\partial\tilde h_{ab}}{\partial t}$\,.

The components of the Riemann tensor can be written in terms of the ADM variables (the so-called Gauss, Codazzi, Ricci--York equations).
We shall only need here the expression of the scalar curvature.
We refer to the literature (see e.g.\ \cite{Misner:1974qy,Wald:1984rg,Gourgoulhon:2007ue}) for its calculation which yields~:
\begin{equation}
\tilde{\mathcal R}
 = {}_\mathbb T\tilde K\,.\,{}_\mathbb T\tilde K
   - \frac{2}{3}\,\tilde K^2
   + \tilde R
   + \frac{2}{\sqrt{-\tilde g}}
     \partial_i (\sqrt{-\tilde g}\,\tilde n^i\,\tilde K)
   - \frac{2}{\sqrt{\tilde h}\,\tilde N}
     \partial_a (\sqrt{\tilde h}\,\tilde h^{ab}\,\partial_b\tilde N)\,,
\label{eq:ERY}
\end{equation}
where $\tilde K\equiv \tilde h^{ab}\tilde K_{ab}$~;
where we place the symbol $\mathbb T$ in front of symmetric tensors to mean their traceless part, e.g.~: ${}_\mathbb TK_{ab}\equiv\tilde K_{ab}-\frac{1}{3}\,\tilde h_{ab}\,\tilde K$~;
where $\tilde K\,.\,\tilde K\equiv\tilde K_{ab}\,\tilde K^{ab}$~;
and where $\tilde R$ is the scalar curvature of the metric $\tilde h_{ab}$\,.
\\
 
Armed with these standard preliminaries consider now the Einstein--scalar action
\begin{equation}
\tilde S_\mathrm E[\tilde g_{ij},\tilde\phi]
 = \int_\mathcal M\!\mathrm d^4x\,\sqrt{-\tilde g}\,
   \left[
    \frac{\tilde{\mathcal R}}{2}
    - \frac{1}{2}\,\tilde g^{ij}\,\partial_i\tilde\phi\,\partial_j\tilde\phi
    - V(\tilde\phi)
   \right]\,.
\label{eq:Eaction}
\end{equation}
This action describes $f(\mathcal R)$ gravity in the ``Einstein frame'' if the potential $V(\tilde\phi)$ is given under parametric form by~:
\begin{equation}
V(s)
 = \frac{s\,f'(s) - f(s)}{2\,f'(s)^2}\,,
\quad
\tilde\phi(s)
 = \sqrt\frac{3}{2}\,\ln f'(s)\,,
\label{eq:Edefphi}
\end{equation}
where a prime denotes derivation with respect to the argument.
As for the metric $\tilde g_{ij}$ it is related to the original, ``Jordan frame'' metric $g_{ij}$ by
\begin{equation}
\tilde g_{ij}
 = e^{\sqrt\frac{2}{3}\tilde\phi}\,g_{ij}\,,
\label{eq:Edefconf}
\end{equation}
see Appendix~\ref{sec:J2E} for a recap.
Following the standard procedure we plug (\ref{eq:ERY}) into (\ref{eq:Eaction}) to get
\begin{equation}
\tilde S_\mathrm E
 = \int_\mathcal M\!\mathrm d^4x\,
   \left[
    \tilde{\mathcal L}_\mathrm E
    + \partial_i (\sqrt{-\tilde g}\,\tilde n^i\,\tilde K)
    - \partial_a (\sqrt{\tilde h}\,\tilde h^{ab}\,\partial_b\tilde N)
   \right]
\label{eq:Eaction3+1}
\end{equation}
with
\begin{equation}
\tilde{\mathcal L}_\mathrm E[\tilde h_{ab},\tilde\phi,\tilde N,\tilde N^a]
 = \sqrt{\tilde h}\tilde N
   \left[
    \frac{1}{2}
    \left(
     {}_\mathbb T\tilde K\,.\,{}_\mathbb T\tilde K
     - \frac{2}{3}\,\tilde K^2
     + \tilde R
    \right)
    + \frac{1}{2\,\tilde N^2}\,
      (\dot{\tilde\phi} - \tilde N^a \partial_a \tilde\phi)^2 
    - \frac{1}{2}\,\partial_a\tilde\phi\,\tilde\partial^a\tilde\phi
    - V(\tilde\phi)
   \right]\,.
\end{equation}

Let us now turn to the obtention of the ADM Hamiltonian \cite{Arnowitt:1962hi}.
Momenta conjugate to the dynamical variables $ \tilde h_{ab} $ and $ \tilde\phi $ are defined as (recalling the definition (\ref{eq:EdefK}) of $\tilde K_{ab}$)
\begin{equation}
\tilde p^{ab}
 \equiv
   \frac{\partial\tilde{\mathcal L}_\mathrm E}{\partial\dot{\tilde h}_{ab}}
 = \frac{\sqrt{\tilde h}}{2}
   \left(
    {}_\mathbb T\tilde K^{ab}
    - \frac{2}{3}\,\tilde K\,\tilde h^{ab}
   \right)\,,
\quad
\tilde\pi
 \equiv
   \frac{\partial\tilde{\mathcal L}_\mathrm E}{\partial\dot{\tilde\phi}}
 = \frac{\sqrt{\tilde h}}{\tilde N}\,
   (\dot{\tilde\phi} - \tilde N^a \partial_a \tilde\phi)\,.
\label{eq:Edefmom}
\end{equation}
Inversion yields the ``velocities'' in terms of the canonical variables~:
\begin{equation}
\dot{\tilde h}_{ab}
 = \frac{4 \tilde N}{\sqrt{\tilde h}}
   \left(
    \tilde p_{ab}
    - \frac{1}{2}\,\tilde p\,\tilde h_{ab}
   \right)
   + \tilde D_a \tilde N_b + \tilde D_b \tilde N_a\,,
\quad
\dot{\tilde\phi}
 = \frac{\tilde N}{\sqrt{\tilde h}}\,\tilde\pi
   + \tilde N^a \partial_a \tilde\phi\,,
\end{equation}
where $\tilde p\equiv\tilde h_{ab}\,\tilde p^{ab}$\,.
Ignoring the divergences in (\ref{eq:Eaction3+1}) the Hamiltonian density is therefore
\begin{equation}
\tilde{\mathcal H}
 \equiv
   \tilde p^{ab}\,\dot{\tilde h}_{ab}
   + \tilde\pi\,\dot{\tilde\phi}
   - \tilde{\mathcal L}_\mathrm E
 = \tilde{\mathcal H}_\mathrm E
   + \partial_a (2\,\tilde p^{ab}\,\tilde N_b)\,,
\quad\hbox{where}\quad
\tilde{\mathcal H}_\mathrm E
 = \sqrt{\tilde h}\,(\tilde N\,\tilde C + \tilde N^a\,\tilde C_a)
\label{eq:EdefHam}
\end{equation}
and
\begin{equation}
\left\{
\begin{aligned}
\tilde C
 & = \frac{2}{\tilde h}
     \left(
      {}_\mathbb T\tilde p\,.\,{}_\mathbb T\tilde p
      - \frac{1}{6}\,\tilde p^2
      + \frac{\tilde\pi^2}{4}
     \right)
     - \frac{\tilde R}{2}
     + \frac{1}{2}\,\partial_a\tilde\phi\,\tilde\partial^a\tilde\phi
     + V(\tilde\phi)\,, \\
\tilde C_a
 & = -2 \tilde D_b\,\left(\frac{\tilde p^b_a}{\sqrt{\tilde h}}\right)
     + \frac{\tilde\pi}{\sqrt{\tilde h}}\, \partial_a \tilde\phi\,.
\end{aligned}
\right.
\label{eq:Edefcons}
\end{equation}
As first shown in \cite{Arnowitt:1962hi} Hamilton's equations
\begin{equation}
\left\{
\begin{aligned}
 &
\tilde C
 = 0\,,
\quad
\tilde C_a
 = 0\,, \\
 &
\frac{\delta\tilde{\mathcal H}_\mathrm E}{\delta\tilde p^{ab}}
 = \dot{\tilde h}_{ab}\,,
\quad
\frac{\delta\tilde{\mathcal H}_\mathrm E}{\delta\tilde h_{ab}}
 = -\dot{\tilde p}^{ab}\,,
\quad
\frac{\delta\tilde{\mathcal H}_\mathrm E}{\delta\tilde\pi}
 = \dot{\tilde\phi}\,,
\quad
\frac{\delta\tilde{\mathcal H}_\mathrm E}{\delta\tilde\phi}
 = -\dot{\tilde\pi}
\end{aligned}
\right.
\end{equation}
are equivalent to Einstein's equations $\tilde G_{ij}=\partial_i\tilde\phi\partial_j\tilde\phi-\tilde g_{ij}\left(\frac{1}{2}(\tilde\partial\tilde\phi)^2+V(\tilde\phi)\right)$\,.

\section{\label{sec:J}
Jordan frame Hamiltonian of $ f(\mathcal R) $ gravity
}

Consider now $ f(\mathcal R) $ gravity in its original ``Jordan frame'' formulation.
The action is
\begin{equation}
S[g_{ij}]
 = \frac{1}{2}
   \int_\mathcal M\!\mathrm d^4x\,\sqrt{-g}\,
   f(\mathcal R)\,.
\end{equation}
The form of the equations of motion (see Appendix~\ref{sec:J2E}) suggests to promote the scalar curvature $\mathcal R$ to the status of independent variable \cite{Schmidt:1994iz}, $s$\,.
We are thus led to replace $S[g_{ij}]$ by the extended action
\begin{equation}
S_\mathrm S[g_{ij},s,\phi]
 = \frac{1}{2}
   \int_\mathcal M\!\mathrm d^4x\,\sqrt{-g}\,
   [f(s) - \phi\, (s - \mathcal R)]\,,
\label{eq:Jaction}
\end{equation}
where $\phi$ is a Lagrange multiplier.
As in Section~\ref{sec:E}, the metric, see (\ref{eq:EADM}), and $\mathcal R$\,, see (\ref{eq:ERY}), are now expressed in terms of the ADM variables as
\begin{equation}
g_{00}
 = -N^2 + N_a N^a\,,
\quad
g_{0a}
 = N_a\,,
\quad
g_{ab}
 = h_{ab}
\label{eq:JADM}
\end{equation}
and
\begin{equation}
\mathcal R
 = {}_\mathbb TK\,.\,{}_\mathbb TK
   - \frac{2}{3}\,K^2
   + R
   + \frac{2}{\sqrt{-g}} \partial_i (\sqrt{-g}\,n^i\,K)
   - \frac{2}{\sqrt h\,N} \partial_a (\sqrt h\,h^{ab} \partial_b N)\,,
\label{eq:JRY}
\end{equation}
where the induced metric on the surface $\Sigma_t$\,, the lapse and the shift are denoted by $\{h_{ab},N,N^a\} $\,, where $n^i$ is the unit vector orthogonal to $\Sigma_t$\,, where $ D_a $ is the covariant derivative associated with $ h_{ab} $\,, where $R$ is the scalar curvature of $\Sigma_t$ and where $\sqrt{-g}=N\,\sqrt h$\,.
Finally, $K_{ab}$ is the extrinsic curvature~:
\begin{equation}
K_{ab}
 = \frac{1}{2\,N}\,(\dot h_{ab} - D_a N_b - D_b N_a)\,.
\label{eq:JdefK}
\end{equation}
Plugging (\ref{eq:JRY}) into (\ref{eq:Jaction}) we have, after integrations by part
\begin{equation}
S_\mathrm S
 = \int_{\mathcal M}\!\mathrm d^4x\,
   \left[
    \mathcal L_\mathrm J
    + \partial_i (\sqrt{-g}\,\phi\,K\,n^i)
    - \partial_a (\sqrt h\,\phi\,h^{ab}\,\partial_b N)
   \right]
\label{eq:Jaction3+1}
\end{equation}
with
\begin{equation}
\mathcal L_\mathrm J[h_{ab},s,N,N^a,\phi]
 = \sqrt h\,N
   \left[
    \frac{\phi}{2}
    \left(
     {}_\mathbb TK\,.\,{}_\mathbb TK
     - \frac{2}{3}\,K^2
     + R
     - s
    \right)
    + \frac{1}{2}\,f(s)
    - \frac{K}{N}\,
      (\dot\phi - N^a \partial_a \phi)
    + \frac{1}{N}\,\partial_a \phi\,\partial^a N
   \right]\,.
\label{eq:JdefLag}
\end{equation}
We thus see that the integration by parts that we have performed has turned $\phi$ into a dynamical field since its time derivative appears in (\ref{eq:JdefLag}).
Now, the equation of motion for $s$ simply is
\begin{equation}
f'(s) = \phi\,.
\label{eq:Jdefs}
\end{equation}
This algebraic constraint can harmlessly be incorporated in $\mathcal L_\mathrm J$ (at least at the classical level) so that the Lagrangian density of the theory becomes\footnote{The divergences in (\ref{eq:Jaction3+1}) are discarded. For a thorough discussion of boundary terms in $f(\mathcal R)$ gravity, see \cite{Dyer:2008hb}.} 
\begin{equation}
\mathcal L^*_\mathrm J[h_{ab},\phi,N,N^a]
 = \mathcal L_\mathrm J[h_{ab},s,N,N^a,\phi]\,,
\end{equation}
where $s$ is known in terms of $\phi$ via (\ref{eq:Jdefs})\,.
(Note that we could have followed an alternative route consisting in {\it first} incorporating the constraint (\ref{eq:Jdefs}) in (\ref{eq:Jaction}) to eliminate $\phi$ and {\it then} turning $s$ into a dynamical variable, the ``scalaron'' \cite{Starobinsky:1980te}.
See Appendix~\ref{sec:toy} and \cite{Deruelle:prep,Sendouda:prep} for a comparison of these two routes.)
\\

Momenta conjugate to the dynamical variables $h_{ab}$ and $\phi$ are defined as, recalling the definition (\ref{eq:JdefK}) of $K_{ab}$~:
\begin{equation}
p^{ab}
 \equiv
   \frac{\partial\mathcal L^*_\mathrm J}{\partial\dot h_{ab}}
 = \frac{\sqrt h}{2}
   \left[
    \phi
    \left({}_\mathbb TK^{ab} - \frac{2}{3}\,K\,h^{ab}\right)
    - \frac{h^{ab}}{N}\,
      (\dot\phi - N^a \partial_a \phi)
   \right]\,,
\quad
\pi
 \equiv
   \frac{\partial\mathcal L^*_\mathrm J}{\partial\dot\phi}
 = -\sqrt h\,K\,.
\label{eq:Jdefmom}
\end{equation}
Inversion yields the velocities in terms of the canonical variables~:
\begin{equation}
\dot h_{ab}
 = \frac{N}{\sqrt h}
   \left(
    \frac{4\,{}_\mathbb Tp_{ab}}{\phi}
    - \frac{2}{3}\,\pi\,h_{ab}
   \right)
   + D_a N_b + D_b N_a\,,
\quad
\dot\phi
 = \frac{2\,N}{3\,\sqrt h}\,(\phi\,\pi - p)
   + N^a \partial_a \phi\,,
\end{equation}
where $p\equiv h_{ab}\, p^{ab}$\,.
The Hamiltonian density is therefore 
\begin{equation}
\mathcal H
 \equiv
   p^{ab}\,\dot h_{ab} + \pi\,\dot\phi - \mathcal L_\mathrm J^*
 = \mathcal H^*_\mathrm J
   + \partial_a (2\,p^{ab}\,N_b - \sqrt h\,\phi\,\partial^a N)\,,
\quad\hbox{where}\quad
\mathcal H^*_\mathrm J
 = \sqrt h\,(N\,C + N^a\,C_a)
\label{eq:JdefHam}
\end{equation}
and
\begin{equation}
\left\{
\begin{aligned}
C
 & = \frac{2}{h}
     \left(
      \frac{{}_\mathbb Tp\,.\,{}_\mathbb Tp}{\phi}
      + \frac{1}{6}\,\phi\,\pi^2
      - \frac{1}{3}\,p\,\pi
     \right)
     + \frac{1}{2}\,
       (\phi\,s - f(s) - \phi\,\bar R + 2 D_a D^a \phi)\,, \\
C_a
 & = -2 D_b\left(\frac{p^b_a}{\sqrt h}\right)
     + \frac{\pi}{\sqrt h}\,\partial_a \phi\,,
\end{aligned}
\right.
\label{eq:Jdefcons}
\end{equation}
where $s$ is known via $f'(s)=\phi$\,.

\section{\label{sec:O}
Ostrogradsky Hamiltonian of $f(\mathcal R)$ gravity
}

Let us return to the $ f(\mathcal R) $ Jordan frame action 
\begin{equation}
S[g_{ij}]
 = \frac{1}{2}
   \int_\mathcal M\!\mathrm d^4x\,\sqrt{-g}\,f(\mathcal R)
\end{equation}
and, contrarily to what we did in the previous section, let us perform the ADM decomposition first, before introducing any new independent variable.

As in Section~\ref{sec:J}, see (\ref{eq:JRY}), $\mathcal R$ is expressed in terms of the ADM variables as
\begin{equation}
\mathcal R
 = {}_\mathbb TK\,.\,{}_\mathbb TK
   - \frac{2}{3}\,K^2
   + R
   + 2 \nabla_i (n^i\, K)
   - \frac{2}{N} D_a D^a N
\end{equation}
that we rewrite as\footnote{Using the relation $\nabla_i n^i=K$ which follows from the preliminaries of Section~\ref{sec:E}.}
\begin{equation}
\mathcal R
 = \frac{2}{N}\,(\dot K - N^a \partial_a K)
   + {}_\mathbb TK\,.\,{}_\mathbb TK
   + \frac{4}{3}\,K^2
   + R
   - \frac{2}{N} D_a D^a N\,.
\label{eq:ORY}
\end{equation}
We recall too the definition of the extrinsic curvature~:
\begin{equation}
K
 = \frac{1}{2\,N}\,(h^{ab}\,\dot h_{ab} - 2\,D_a N^a)\,,
\quad
{}_\mathbb TK_{ab}
 = \frac{1}{2\,N}\,[{}_\mathbb T\dot h_{ab} - {}_\mathbb T(D_a N_b+ D_b N_a)]\,.
\label{eq:OdefK}
\end{equation}
We hence see explicitly that the scalar curvature depends on second time derivatives of $h_{ab}$ through $\dot K$\,.
This suggests \cite{Buchbinder:1987vp} to promote, ``\`a la Ostrogradsky,'' $K$ to the status of a new independent variable, $Q$ (see also \cite{Querella:1998ke,Ezawa:1998ax}).
We are thus led to replace $S[g_{ij}]$ by the extended action
\begin{equation}
S_\mathrm O
 = \int_\mathcal M\!\mathrm d^4x\,\mathcal L_\mathrm O\,,
\quad\hbox{where}\quad
\mathcal L_\mathrm O[h_{ab},Q,N,N^a,u]
 = \sqrt h\,N\,
   \left[\frac{1}{2}\,f(\mathcal R) + u\,(K - Q)\right]
\end{equation}
with $\mathcal R$ now given as
\begin{equation}
\mathcal R
 = \frac{2}{N}\,(\dot Q - N^a \partial_a Q)
   + {}_\mathbb TK\,.\,{}_\mathbb TK
   + \frac{4}{3}\,Q^2
   + R
   - \frac{2}{N} D_a D^a N
\label{eq:ORaux}
\end{equation}
and where $K$ and ${}_\mathbb TK_{ab}$ are given in (\ref{eq:OdefK}).\footnote{Note that we chose to replace $K$ by $Q$ everywhere in (\ref{eq:ORY}), but in the expression (\ref{eq:OdefK}) of ${}_\mathbb TK_{ab}$ in order to keep it traceless. See Appendix~\ref{sec:toy} for examples of alternative choices.}
\\

Momenta conjugate to the dynamical variables $ h_{ab} $ and $ Q$ are defined as~:
\begin{equation}
P^{ab}
 \equiv
   \frac{\partial\mathcal L_\mathrm O}{\partial\dot h_{ab}}
 = \frac{\sqrt h}{2}
   (f'(\mathcal R)\,{}_\mathbb TK^{ab} + u\,h^{ab})\,,
\quad
\Pi
 \equiv
   \frac{\partial\mathcal L_\mathrm O}{\partial\dot Q}
 = \sqrt h\,f'(\mathcal R)\,,
\label{eq:Odefmom}
\end{equation}
where $\mathcal R$ is given in (\ref{eq:ORaux}).
Inversion yields~:
\begin{equation}
\left\{
\begin{aligned}
{}_\mathbb T\dot h_{ab}
 & = \frac{4\,N}{\Pi} {}_\mathbb TP_{ab}
     + {}_\mathbb T(D_a N_b + D_b N_a)\,, \\
\dot Q
 & = \frac{N}{2}
     \left(
      \mathcal R
      - 4 \frac{{}_\mathbb TP\,.\,{}_\mathbb TP}{\Pi^2}
      - \frac{4}{3}\,Q^2
      - R
     \right)
     + D_a D^a N
     + N^a \partial_a Q\,,
\end{aligned}
\right.
\end{equation}
where $\mathcal R$ is known in terms of $\Pi/\sqrt h$ via $f'(\mathcal R)=\Pi/\sqrt h$\,.
The Lagrangian $\mathcal L_\mathrm O$ is therefore singular in that it cannot be inverted to give the trace of the velocities $\dot h_{ab}$\,.
However the Hamiltonian density
\begin{equation}
\mathcal H_\mathrm O
 = P^{ab}\,\dot h_{ab} + \Pi\,\dot Q - \mathcal L_\mathrm O
\end{equation}
is still well defined if one injects in $\mathcal L_\mathrm O$ the constraint stemming from (\ref{eq:Odefmom}), to wit, $u=\frac{2P}{3\sqrt h}$\,.\footnote{See Appendix~\ref{sec:toy} and \ref{sec:mini} for illustrations of the same phenomenon on toy models.}
It reads
\begin{equation}
\mathcal H_\mathrm O
 = \mathcal H_\mathrm O^*
   + \partial_a
     \left[
      \Pi\,\partial^a N
      - \sqrt h\,N\,\partial^a \left(\frac{\Pi}{\sqrt h}\right)
      + 2\,P^{ab}\,N_b
     \right]\,,
\quad\hbox{where}\quad
\mathcal H_\mathrm O^*
 = \sqrt h\,(N\,C^\mathrm O + N^a\,C^\mathrm O_a)
\label{eq:OdefHam}
\end{equation}
and
\begin{equation}
\left\{
\begin{aligned}
C^\mathrm O
 & = \frac{2}{\sqrt h}
     \left(
      \frac{{}_\mathbb TP\,.\,{}_\mathbb TP}{\Pi}
      + \frac{1}{3}\,P\,Q
     \right)
     + \frac{\Pi}{2\sqrt h}
       \left(\mathcal R - R - \frac{4}{3}\,Q^2\right)
     - \frac{1}{2}\,f(\mathcal R)
     + D_a D^a \left(\frac{\Pi}{\sqrt h}\right)\,, \\
C_a^\mathrm O
 & = -2 D_b \left(\frac{\,P^b_a}{\sqrt h}\right)
     + \frac{\Pi}{\sqrt h}\,\partial_a Q\,,
\end{aligned}
\right.
\label{eq:Odefcons}
\end{equation}
where $\mathcal R$ is known via $f'(\mathcal R)=\Pi/\sqrt h$\,.

\section{\label{sec:trans}
Canonical transformations
}

In the previous sections we have associated, as concisely as possible, three seemingly different Hamiltonians to $f(\mathcal R)$ gravity~: the ``Ostrogradsky'' Hamiltonian $\mathcal H_\mathrm O^*$ (\ref{eq:OdefHam}) (\ref{eq:Odefcons}), and two ``Schmidt'' \cite{Schmidt:1994iz} Hamiltonians~: the ``Jordan frame'' one $\mathcal H_\mathrm J^*$ (\ref{eq:JdefHam}) (\ref{eq:Jdefcons}), and the ``Einstein'' one $\tilde{\mathcal H}_\mathrm E$ (\ref{eq:EdefHam}) (\ref{eq:Edefcons}).
Despite some claims to the contrary \cite{Ezawa:1998ax,Ezawa:2009rh}, their respective sets of canonical variables
\begin{equation}
\mathcal H_\mathrm O^*~:
\{h_{ab},P^{ab},Q,\Pi,N,N^a\}\,,
\quad
\mathcal H_\mathrm J^*~:
\{h_{ab},p^{ab},\phi,\pi,N,N^a\}\,,
\quad
\tilde{\mathcal H}_\mathrm E~:
\{\tilde h_{ab},\tilde p^{ab},\tilde\phi,\tilde\pi,\tilde N,\tilde N^a\}
\end{equation}
turn out to be related by means of canonical transformations.
We proceed to show this explicitly.

\subsection{Einstein $\rightarrow$ Jordan}

The Einstein frame metric is conformally related to the Jordan frame one, see (\ref{eq:Edefconf}).
Thus the relation between the ADM variables $\{\tilde h_{ab},\tilde N,\tilde N^a\}$ and $\{h_{ab},N,N^a\}$ is known, see (\ref{eq:EADM}) and (\ref{eq:JADM}).
Taking then into account the relation between $\tilde\phi$\,, $f'(s)$ and $\phi$\,, see (\ref{eq:Edefphi}) and (\ref{eq:Jdefs}), we therefore have\footnote{Note that we must have $\phi>0$ (which is equivalent to requiring that the two metrics be related by a positive conformal factor).}
\begin{equation}
\tilde h_{ab}
 = \phi\,h_{ab}\,,
\quad
\tilde N^a
 = N^a\,,
\quad
\tilde N
 = \sqrt\phi\,N\,,
\quad
\tilde\phi
 = \sqrt\frac{3}{2}\,\ln\phi\,.
\end{equation}
Now, since the extrinsic curvatures of the two frames, see (\ref{eq:EdefK}) and (\ref{eq:JdefK}), are related thus
\begin{equation}
\tilde K_{ab}
 = \sqrt\phi\,K_{ab}
   + \frac{h_{ab}}{2\,N\,\sqrt h}\,(\dot\phi - N^a \partial_a \phi)\,,
\end{equation}
we deduce from (\ref{eq:Edefmom}) and (\ref{eq:Jdefmom}) that the momenta are given by
\begin{equation}
\tilde p^{ab}
 = \frac{1}{\phi}\,p^{ab}\,,
\quad
\tilde\pi
 = \sqrt\frac{2}{3}\,(\phi\,\pi - p)\,.
\end{equation}

If we now plug these expressions of the Einstein variables in the Einstein Hamiltonian $\tilde{\mathcal H}_\mathrm E$ given in (\ref{eq:EdefHam}) (\ref{eq:Edefcons}),we find that $\tilde{\mathcal H}_\mathrm E$ turns into~:\footnote{After developing $\tilde D_a$ in terms of $D_a$ and $\tilde R$ in terms of $R$ and recalling that $V(\tilde\phi)$ is given in terms of $\phi$ via (\ref{eq:Edefphi}) and (\ref{eq:Jdefs}) as~: $V=\frac{s\,\phi-f(s)}{2\,\phi^2}$ with $s$ known via $f'(s)=\phi$\,.}
\begin{equation}
\tilde{\mathcal H}_\mathrm E
\,\longrightarrow\,
\mathcal H^*_\mathrm J\,,
\end{equation}
where $\mathcal H^*_\mathrm J$ is the Jordan Hamiltonian given in (\ref{eq:JdefHam}) (\ref{eq:Jdefcons}).

Moreover the transformation is canonical~:
if the Poisson bracket of two functions $A$ and $B$ of the Jordan variables $\{h_{ab},p^{ab},\phi,\pi,N,N^a\}$\,, is defined as usual by
\begin{equation}
\{A,B\}_\mathrm J
 \equiv
   \frac{\delta A}{\delta p^{ab}} \frac{\delta B}{\delta h_{ab}}
   - \frac{\delta A}{\delta h_{ab}} \frac{\delta B}{\delta p^{ab}}
   + \frac{\delta A}{\delta\pi} \frac{\delta B}{\delta\phi}
   - \frac{\delta A}{\delta\phi} \frac{\delta B}{\delta\pi}
\end{equation}
then it is an exercise to see that (the variational reducing to partial derivatives)~:
$\{\tilde h_{ab},\tilde p^{cd}\}_\mathrm J=\{h_{ab},p^{cd}\}_\mathrm J$\,,
$\{\tilde\phi,\tilde\pi\}_\mathrm J=\{\phi,\pi\}_\mathrm J$\,,
$\{\tilde p^{ab},\tilde\pi\}_\mathrm J=0$\,,
$\{\tilde h_{ab},\tilde\phi\}_\mathrm J=0$\,,
$\{\tilde h_{ab},\tilde\pi\}_\mathrm J=0$\,,
$\{\tilde p^{ab},\tilde\phi\}_\mathrm J=0$\,.
We are therefore guaranteed that $\tilde{\mathcal H}_\mathrm E$ and $\mathcal H^*_\mathrm J$ yield the same equations of motion.
(Since this can be shown separately, see \cite{Sendouda:prep}, the results are watertight.)

\subsection{
Jordan $\rightarrow$ Ostrogradsky 
}

In the Ostrogradsky formulation we introduced as a new variable the extrinsic curvature~: $Q=K$ and found that the scalar curvature $\mathcal R$ was given by $f'(\mathcal R)=\Pi/\sqrt h$\,, $\Pi$ being the momentum conjugate to $Q$\,, see (\ref{eq:Odefmom}).
In the Jordan frame formulation on the other hand the new variable was $s=\mathcal R$ that we traded for $\phi$ via $f'(s)=\phi$\,, see (\ref{eq:Jdefs})~;
as for the extrinsic curvature it was given by $K=-\pi/\sqrt h$\,, $\pi$ being the momentum conjugate to $\phi$\,, see (\ref{eq:Jdefmom}).
All this suggests to choose 
\begin{equation}
\phi
 = \frac{\Pi}{\sqrt h}\,,
\quad
\pi
 = -\sqrt h\, Q\,.
\end{equation} 

Plugging these expressions for $\phi$ and $\pi$ into the Jordan Hamiltonian $\mathcal H^*_\mathrm J$ given in (\ref{eq:JdefHam}) (\ref{eq:Jdefcons}) we find that $H^*_\mathrm J$ identifies to the Ostrogradsky Hamiltonian $\mathcal H^*_\mathrm O$ given in (\ref{eq:OdefHam}) (\ref{eq:Odefcons})~:
\begin{equation}
\mathcal H^*_\mathrm J
\,\longrightarrow\,
\mathcal H^*_\mathrm O
\end{equation}
if we choose
\begin{equation}
p^{ab}
 = P^{ab} - \frac{Q\,\Pi}{2}\,h^{ab}\,.
\end{equation}
We have thus transformed all the Jordan variables $\{h_{ab},p^{ab},\phi,\pi,N,N^a\}$ into the Ostrogradsky ones $ \{h_{ab},P^{ab},Q,\Pi,N,N^a\}$\,.

Again it is easy to see that the transformation is canonical ~:
if the Poisson bracket of two functions $A$ and $B$ of the Ostrogradsky variables $\{h_{ab},P^{ab},Q,\Pi,N,N^a\}$\,, is defined as
\begin{equation}
\{A,B\}_\mathrm O
 \equiv
   \frac{\delta A}{\delta P^{ab}} \frac{\delta B}{\delta h_{ab}}
   - \frac{\delta A}{\delta h_{ab}} \frac{\delta B}{\delta P^{ab}}
   + \frac{\delta A}{\delta\Pi} \frac{\delta B}{\delta Q}
   - \frac{\delta A}{\delta Q} \frac{\delta B}{\delta\Pi}
\end{equation}
then we have~:
$\{h_{ab},p^{cd}\}_\mathrm O=\{h_{ab},P^{cd}\}_\mathrm O$\,,
$\{\phi,\pi\}_\mathrm O=\{Q,\Pi\}_\mathrm O$\,,
$\{h_{ab},\phi\}_\mathrm O=0$\,, $\{h_{ab},\pi\}_\mathrm O=0$\,,
$\{p^{ab},\phi\}_\mathrm O=0$\,, $\{p^{ab},\pi\}_\mathrm O=0$\,.

We are therefore guaranteed that $\tilde{\mathcal H}_\mathrm E$ and $\mathcal
H^*_\mathrm J$ yield the same equations of motion.
(Since we have shown that separately, at least in the minisuperspace case, see
Appendix~\ref{sec:mini}, the results are safe.)

\subsection{
Ostrogradsky $\rightarrow$ Einstein
}

To close the loop we combine the transformations obtained above to get the Ostrogradsky variables in terms of the Einstein ones (note that they are at odds with those advocated in \cite{Ezawa:1998ax} and \cite{Ezawa:2009rh})~:
\begin{equation}
\left\{
\begin{aligned}
 &
h_{ab}
 = e^{-\sqrt\frac{2}{3}\tilde\phi}\,\tilde h_{ab}\,,
\quad
P^{ab}
 = e^{\sqrt\frac{2}{3}\tilde\phi}
   \left[
    \tilde p^{ab}
    - \frac{1}{2}
      \left(\sqrt\frac{3}{2}\,\tilde\pi + \tilde p\right)
      \tilde h^{ab}
   \right]\,, \\
 &
Q
 = -\frac{e^{\tilde\phi/\sqrt6}}{\sqrt{\tilde h}}
   \left(\sqrt\frac{3}{2}\,\tilde\pi + \tilde p\right)\,,
\quad
\Pi
 = \sqrt{\tilde h}\,e^{-\tilde\phi/\sqrt6}\,, \\
 &
N
 = e^{-\tilde\phi/\sqrt6}\,\tilde N\,,
\quad
N^a
 = \tilde N^a\,.
\end{aligned}
\right.
\label{eq:transO2E}
\end{equation}

Plugging these expressions into the Ostrogradsky Hamiltonian $\mathcal H^*_\mathrm O$ given in (\ref{eq:OdefHam}) (\ref{eq:Odefcons}) we find that it transforms into~:\footnote{Again, we have to develop $D_a$ in terms of $\tilde D_a$ and $R$ in terms of $\tilde R$ and recall that $V(\tilde\phi)$ is given in terms of $\phi=e^{\sqrt\frac{2}{3}\tilde\phi}$ via (\ref{eq:Edefphi}) and (\ref{eq:Jdefs}) as~: $V=\frac{s\,\phi-f(s)}{2\,\phi^2}$ with $s$ known via $f'(s)=\phi$\,.}
\begin{equation}
\mathcal H^*_\mathrm O
\,\longrightarrow\,
\tilde{\mathcal H}_\mathrm E\,,
\end{equation}
where the Einstein Hamiltonian $\tilde{\mathcal H}_\mathrm E$ is given in (\ref{eq:EdefHam}) (\ref{eq:Edefcons}).
Since (\ref{eq:transO2E}) is a composition of two canonical transformations it is canonical too.

\section{\label{sec:conc}
Conclusions
}

We have given three seemingly different Hamiltonian formulations of $f(\mathcal R)$ gravity~:
\begin{enumerate}
\item
an ``Einstein frame'' formulation with variables $ \{\tilde h_{ab},\tilde p^{ab},\tilde\phi,\tilde\pi,\tilde N,\tilde N^a\} $\,, where the extra degree of freedom is embodied in the variables $\{\tilde\phi,\tilde\pi\}$ and which is nothing but the ADM formulation of General Relativity minimally coupled to a scalar field,
\item
a ``Jordan frame'' formulation with variables $ \{h_{ab},p^{ab},\phi,\pi,N,N^a\} $\,, where the extra degree of freedom is taken to be the scalar curvature $\mathcal R$ and is represented by the variables $\{\phi,\pi\}$\,,
\item
an ``Ostrogradsky'' formulation with variables $ \{h_{ab},P^{ab},Q,\Pi,N,N^a\} $\,, where the extra degree of freedom is taken to be the extrinsic curvature $K$ and is represented by the variables $\{Q,\Pi\}$\,,
\end{enumerate}
and we have shown that they are all (classically) equivalent since the three sets of variables are related by canonical transformations.

Now these canonical transformations, see e.g.\ (\ref{eq:transO2E}), are highly non-linear.
These theories are therefore unlikely to be equivalent at the quantum level, see e.g.\ \cite{Anderson:1993ia}.
We leave these developments to further work.

\begin{acknowledgments}
We thank Misao Sasaki, Marc Henneaux, Philippe Spindel and Alexei Starobinsky for discussions.
N.D.\ thanks the Yukawa Institute, where this work began, for enduring hospitality. 
Y.S.\ thanks the hospitality of APC, where this work was completed.
Y.S.\ is supported in part by MEXT through a Grant-in-Aid for JSPS Fellows.
\end{acknowledgments}

\appendix
\section{\label{sec:toy}
Hamiltonian formulations of higher derivative theories~: a toy model
}

We gather here some results, most of them already known \cite{Schmidt:1994iz,Buchbinder:1987vp,Woodard:2006nt,Querella:1998ke,Andrzejewski:2007by,Govaerts:1994zh,Nakamura:1995qz}, concerning the following higher derivative action~:
\begin{equation}
S[q]
 = \int_{t_1}^{t_2}\!\mathrm dt\,L
\quad\hbox{with}\quad
L
 = \frac{1}{2}\,\dot q^2 - \frac{1}{2}\,q^2 + g(\ddot q)\,,
\label{eq:toyact}
\end{equation}
where a dot denotes a derivative with respect to time $t$ and where $g$ is an arbitrary function.

Extremisation of $S$ with respect to path variations $\delta q(t)$ such that $\delta q$ and $\delta\dot q$ vanish at the boundaries $t_1$ and $t_2$ yields a fourth order differential Euler--Lagrange equation
\begin{equation}
q + \ddot q - \ddot g' = 0\,,
\label{eq:toyeom}
\end{equation}
where a prime denotes a derivative with respect to the argument.

Since both $\delta q$ and $\delta\dot q$ have to vanish at the boundaries, Ostrogradsky (see e.g.\ \cite{Woodard:2006nt}) suggested to promote
\begin{equation}
Q \equiv \dot q
\label{eq:toydefQ}
\end{equation}
to the status of an independent variable.
The action is thus extended \cite{Querella:1998ke,Govaerts:1994zh,Nakamura:1995qz} to take account of this constraint~: $S\to S_\mathrm O$ with
\begin{equation}
S_\mathrm O[q,Q,u]
 = \int_{t_1}^{t_2}\!\mathrm dt\,L_\mathrm O
\quad\hbox{and}\quad
L_\mathrm O
 = \frac{1}{2}\,Q^2 - \frac{1}{2}\,q^2 + g(\dot Q) + u\,(\dot q - Q)\,,
\end{equation}
where $u$ is a Lagrange multiplier.\footnote{We could as well have extended $L$ into $L_\mathrm O=\frac{1}{2}\dot q^2-\frac{1}{2}q^2+g(\dot Q)+u\,(\dot q-Q)$ or $L_\mathrm O=\frac{1}{2}Q\,\dot q-\frac{1}{2}q^2+g(\dot Q)+u\,(\dot q-Q)$\,. It is easy to see that the respective Hamiltonians all lead to the same equation of motion (\ref{eq:toyeom}).}
Extremisation of $S_\mathrm O$ with respect to $u$\,, $q$ and $Q$ gives $\dot q=Q$\,, $\dot u=-q$ and $\dot g'=Q-u$\,, that is, the equation of motion (\ref{eq:toyeom}).
It is then straightforward to obtain the Hamiltonian $H_\mathrm O$ associated to $L_\mathrm O$\,.
Indeed, the momenta are
\begin{equation}
P
 \equiv
   \frac{\partial L_\mathrm O}{\partial\dot q}
 = u\,,
\quad
\Pi
 \equiv
   \frac{\partial L_\mathrm O}{\partial\dot Q}
 = g'(\dot Q)\,.
\label{eq:toyOdefmom}
\end{equation}
$L_\mathrm O$ is singular in that (\ref{eq:toyOdefmom}) cannot be inverted to give $\dot q$\,.\footnote{The same happens when treating in the same manner the minisuperspace version of $f(\mathcal R)$ gravity, see Appendix~\ref{sec:mini}. The same happens too in the full-fledged version of the theory, where only the traceless part of the velocities can be explicitly expressed in terms of the variables and their momenta, see Section~\ref{sec:O}.}
The Hamiltonian $H^*_\mathrm O\equiv P\,\dot q+\Pi\,\dot Q-L_\mathrm O$ is however still well defined if one injects the constraint $u=P$ in $L_\mathrm O$~:
\begin{equation}
H^*_\mathrm O(q,P,Q,\Pi)
 = \Pi\,\dot Q - \frac{1}{2}\,Q^2 + \frac{1}{2}\,q^2 - g(\dot Q) + P\,Q\,,
\label{eq:toyOdefHam}
\end{equation}
where $\dot Q$ is known in terms of $\Pi$ via $g'(\dot Q)=\Pi$\,, see \cite{Andrzejewski:2007by}.
One checks that the Hamilton equations $\frac{\partial H^*_\mathrm O}{\partial P}=\dot q$\,, $\frac{\partial H^*_\mathrm O}{\partial\Pi}=\dot Q$\,, $\frac{\partial H^*_\mathrm O}{\partial q}=-\dot P$ and $\frac{\partial H^*_\mathrm O}{\partial Q}=-\dot\Pi$ give back (\ref{eq:toyeom}).
In \cite{Buchbinder:1987vp} Buchbinder and Lyahovich showed that it was indifferent to choose $\dot q$ or any function of $q$ and $\dot q$ as the new independent variable since the respective sets of canonical variables are related by canonical transformations.
\\

Now, seemingly different Hamiltonians can be built from (\ref{eq:toyact}) if one decides to promote 
\begin{equation}
s \equiv \ddot q
\label{eq:toydefs}
\end{equation}
rather than $\dot q$\,, as an independent variable \cite{Schmidt:1994iz,Andrzejewski:2007by}.
The action is again extended to take account of the constraint~: $S\to S_\mathrm S$ with
\begin{equation}
S_\mathrm S[q,s,\phi]
 = \int_{t_1}^{t_2}\!\mathrm dt\,L_\mathrm S
\quad\hbox{and}\quad
L_\mathrm S
 = \frac{1}{2}\,\dot q^2
   - \frac{1}{2}\,q^2
   + g(s)
   + \phi\,(\ddot q - s)\,,
\label{eq:toySactaux}
\end{equation}
where $\phi$ is a Lagrange multiplier.
Extremisation of $S_\mathrm S$ with respect to $s$\,, $\phi$ and $q$ gives $\phi=g'(s)$\,, $s=\ddot q$ and $-q=\ddot q-\ddot\phi$\,, that is, (\ref{eq:toyeom}).
\\

The traditional route is, first, to plug the constraint $\phi=g'(s)$ into (\ref{eq:toySactaux}).\footnote{When this is done in the context of $f(\mathcal R)$ gravity one gets the ``Jordan frame'' action, see e.g.\ Appendix~\ref{sec:mini}.}
Pursuing this path means replacing $S_\mathrm S[q,s,\phi]$ by 
\begin{equation}
S_\mathrm{JF}[q,s]
 = \int_{t_1}^{t_2}\!\mathrm dt\,L_\mathrm{JF}
\quad\hbox{with}\quad
L_\mathrm{JF}
 = \frac{1}{2}\,\dot q^2
   - \frac{1}{2}\,q^2
   + g(s)
   + g'(s)\,(\ddot q-s)\,.
\end{equation}
A second step is to add to $S_\mathrm{JF}$ the boundary term $-(g'(s)\dot q)^{t_1}_{t_2}$ and consider~:\footnote{This is our toy model analogue of the Hawking--Luttrell boundary term \cite{Hawking:1984ph}.} 
\begin{equation}
S^*_\mathrm{JF}[q,s]
 = \int_{t_1}^{t_2}\!\mathrm dt\,L^*_\mathrm{JF}
\quad\hbox{with}\quad
L^*_\mathrm{JF}
 = L_\mathrm{JF} - \frac{\mathrm d}{\mathrm dt}(g'(s)\dot q)
 = \frac{1}{2}\,\dot q^2
   - \frac{1}{2}\,q^2
   + g(s)
   - s\,g'(s)
   - g''\,\dot q\,\dot s\,.
\end{equation}
This operation transforms the action into an ordinary one, since $\ddot q$ has disappeared, and, in doing so, turns $s$ into a dynamical variable, since $\dot s$ now appears, albeit only linearly.
(One can check that extremisation of $S_\mathrm{JF}$ and $S^*_\mathrm{JF}$ yields back (\ref{eq:toyeom}).)

The conjugate momenta of $q$ and $s$ are~:
\begin{equation}
\pi_q
 = \frac{\delta S^*_\mathrm{JF}}{\delta\dot q}
 = \dot q - g''\,\dot s\,,
\quad
\pi_s
 = \frac{\delta S^*_\mathrm{JF}}{\delta\dot s}
 = -g''\,\dot q\,.
\label{eq:toyJFdefmom}
\end{equation}
Inversion of (\ref{eq:toyJFdefmom}) is possible only if $g$ is non-linear.
The Hamiltonian $H^*_\mathrm{JF}\equiv \pi_q\,\dot q+\pi_s\,\dot s-L^*_\mathrm{JF}$ then is \cite{Andrzejewski:2007by}
\begin{equation}
H^*_\mathrm{JF}(q,\pi_q,s,\pi_s)
 = -\frac{1}{2}\,\frac{\pi_s^2}{g''^2}
   - \frac{\pi_s\,\pi_q}{g''}
   - g(s)
   + s\,g'
   + \frac{1}{2}\,q^2\,.
\label{eq:toyJFdefHam}
\end{equation}
One can check that the Hamilton equations give back (\ref{eq:toyeom}).
Since $H^*_\mathrm{JF}$ is singular when $g$ is linear we prefer to keep $s$ and $\phi$ as independent variables in (\ref{eq:toySactaux}).
\\
 
Returning then to (\ref{eq:toySactaux}) we {\sl first} eliminate the $\ddot q$ term by adding the boundary term $-(\dot q\phi)^{t_2}_{t_1}$ and consider, instead of $S_\mathrm S$~:\footnote{We thank Misao Sasaki for discussing with us this alternative procedure.}
\begin{equation}
S_\mathrm J[q,s,\phi]
 = \int_{t_1}^{t_2}\!\mathrm dt\,L_\mathrm J
\quad\hbox{with}\quad
L_\mathrm J
 = L_\mathrm S - \frac{\mathrm d}{\mathrm dt}(\dot q\,\phi)
 = \frac{1}{2}\,\dot q^2
   - \frac{1}{2}\,q^2
   + g(s)
   - \phi\,s
   - \dot q\,\dot\phi\,,
\label{eq:toyJactaux}
\end{equation}
where now $\phi$ is a dynamical variable.\footnote{See Section~\ref{sec:J} and \cite{Deruelle:prep,Sendouda:prep} for application to $f(\mathcal R)$ gravity.}
Extremisation of $S_\mathrm J$ with respect to $s$\,, $\phi$ and $q$ gives, as before, $g'(s)=\phi$\,, $s=\ddot q$ and $-q=\ddot q-\ddot\phi$\,, that is, (\ref{eq:toyeom}).
It is at this stage that we plug the constraint $g'(s)=\phi$ into (\ref{eq:toyJactaux}) and replace $S_\mathrm J$ by
\begin{equation}
S^*_\mathrm J[q,\phi]
 = \int_{t_1}^{t_2}\!\mathrm dt\,L^*_\mathrm J
\quad\hbox{with}\quad
L^*_\mathrm J
 = \frac{1}{2}\,\dot q^2
   - \frac{1}{2}\,q^2
   + g(s)
   - \phi\,s
   - \dot q\,\dot\phi\,,
\label{eq:toyJact}
\end{equation}
where $s$ is known via $g'(s)=\phi$\,.

The conjugate momenta of $q$ and $\phi$ are
\begin{equation}
p
 \equiv
   \frac{\delta S^*_\mathrm J}{\delta\dot q}
 = \dot q - \dot\phi\,,
\quad
\pi
 \equiv
   \frac{\delta S^*_\mathrm J}{\delta\dot\phi}
 = -\dot q\,.
\label{eq:toyJdefmom}
\end{equation}
Contrarily to (\ref{eq:toyJFdefmom}) these momenta are invertible even if $g$ is linear and the Hamiltonian reads
\begin{equation}
H^*_\mathrm J(q,p,\phi,\pi)
 = -\frac{1}{2}\,\pi^2
   - \pi\,p
   - g(s)
   + s\,\phi
   + \frac{1}{2}\,q^2\,,
\label{eq:toyJdefHam}
\end{equation}
where $s$ is known in terms of $\phi$ via $g'(s)=\phi$\,.
The limit $g=s$ is obtained by ``freezing the extra degree of freedom,'' that is, setting $\phi=1$\,, either in the action (\ref{eq:toyJact}), or in (\ref{eq:toyJdefmom}) which then gives $p=-\pi$ so that the Hamiltonian (\ref{eq:toyJdefHam}) reduces to~: $H^*_\mathrm J=\frac{1}{2}p^2+\frac{1}{2}q^2$\,.
\\

We have thus associated three seemingly different Hamiltonians to the original action (\ref{eq:toyact})~:
the ``Ostrogradsky'' Hamiltonian $H_\mathrm O^*$ (\ref{eq:toyOdefHam}), and two ``Schmidt'' Hamiltonians~: $H^*_\mathrm{JF}$ (\ref{eq:toyJFdefHam}), and $H^*_\mathrm J$ (\ref{eq:toyJdefHam}).
Since they all yield the same equations of motion (\ref{eq:toyeom}) it should not come as a surprise that their respective sets of canonical variables, to wit
\begin{equation}
H^*_\mathrm J~: \{q,p,\phi,\pi\}\,,
\quad
H^*_\mathrm{JF}~: \{q,\pi_q,s,\pi_s\}\,,
\quad
H^*_\mathrm O~: \{q,P,Q,\Pi\}
\end{equation}
are related by means of canonical transformations.

Let us start with the correspondence $H^*_\mathrm J\to H^*_\mathrm{JF}$\,.
Equations (\ref{eq:toyJFdefmom}) and (\ref{eq:toyJdefmom}) suggest to choose
\begin{equation}
\phi
 = g'(s)
\quad\hbox{and}\quad
\pi
 = \frac{\pi_s}{g''(s)}\,.
\end{equation}
Plugging these expressions into (\ref{eq:toyJdefHam}) gives $H^*_\mathrm J=H^*_\mathrm{JF}$ if
\begin{equation}
p = \pi_q
\end{equation}
and it is an exercise to check that the transformation is canonical ~:
indeed, the Poisson brackets of the set $\{A,B\}=\{q,p,\phi,\pi\}$ with respect to the set $ \{q,\pi_q,s,\pi_s\}$ being defined as~:
\begin{equation}
\{A,B\}_\mathrm{JF}
 \equiv
   \frac{\partial A}{\partial\pi_s} \frac{\partial B}{\partial s}
   - \frac{\partial A}{\partial s} \frac{\partial B}{\partial\pi_s}
   + \frac{\partial A}{\partial\pi_q} \frac{\partial B}{\partial q}
   - \frac{\partial A}{\partial q} \frac{\partial B}{\partial\pi_q}
\end{equation}
are canonical, that is~:
$\{q,p\}_\mathrm{JF}=-1$\,,
$\{\phi,\pi\}_\mathrm{JF}=-1$\,,
$\{q,\phi\}_\mathrm{JF}=0$\,,
$\{q,\pi\}_\mathrm{JF}=0$\,,
$\{p,\phi\}_\mathrm{JF}=0$\,,
$\{p,\pi\}_\mathrm{JF}=0$\,.
\\

Let us now turn to the correspondence $H^*_\mathrm{JF}\to H^*_\mathrm O$ \cite{Andrzejewski:2007by}.
Equations (\ref{eq:toydefQ}) (\ref{eq:toyOdefmom}) (\ref{eq:toydefs}) and (\ref{eq:toyJFdefmom}) suggest to choose
\begin{equation}
\pi_s
 = -g''(s)\,Q\,,
\end{equation}
where $s$ is known in terms of $\Pi$ via $g'(s)=\Pi$\,.
Plugging these expressions into (\ref{eq:toyJFdefHam}) gives $H^*_\mathrm{JF}=H^*_\mathrm O$ if
\begin{equation}
\pi_q=P
\end{equation}
after renaming the parameter $s$ as $s=\dot Q$\,.
Again it is an exercise to compute the Poisson brackets of the set $\{q,\pi_q,s,\pi_s\}$ with respect to the set $\{q, P, Q,\Pi\}$ and see that the transformation is canonical.
\\

We have thus shown (in full details) the canonical equivalence of three different Hamiltonian formulations of our toy model, akin to those employed when treating $f(\mathcal R)$ gravity.

\section{\label{sec:J2E}
From the Jordan to the Einstein frame~: a short recap
}

We recall here how the action for $f(\mathcal R)$ gravity is transformed into the Hilbert action for a conformally rescaled metric minimally coupled to a scalar field \cite{Teyssandier:1983zz}.

Consider the action for $f(\mathcal R)$ gravity~: 
\begin{equation}
S[g_{ij}]
 = \frac{1}{2} \int\!\mathrm d^4x\,\sqrt{-g}\,f(\mathcal R)
   + S_\mathrm m[\Psi,g_{ij}]\,,
\label{eq:f(R)act}
\end{equation}
where Einstein's constant $\kappa\equiv8\pi G=1$\,, where $g$ is the determinant of the metric $g_{ij}$ with signature $(-,+,+,+)$\,, where $\mathcal R=\frac{1}{2}(g^{ik}\,g^{jl}-g^{ij}\,g^{kl})\partial_{ij}g_{kl}+\cdots$ is the scalar curvature, and where $\Psi$ denotes some matter fields minimally coupled to the metric.
Since (\ref{eq:f(R)act}) contains second derivatives of $g_{ij}$ which do not sum up as a divergence (unless $f=\mathcal R$) its extremisation with respect to metric variations yields fourth-order differential field equations~:
\begin{equation}
f'\,G_{ij}
+ \frac{1}{2}\,g_{ij}\,(\mathcal R\,f' - f)
- D_{ij} f'
+ g_{ij}\,\square f'
 = T^\mathrm m_{ij}\,,
\label{eq:f(R)eom}
\end{equation}
where a prime denotes a derivative with respect to the argument, where $G_{ij}$ is Einstein's tensor and where $T^\mathrm m_{ij}=-\frac{2}{\sqrt{-g}} \frac{\delta S_\mathrm m}{\delta g^{ij}}$ is the matter stress-energy tensor.
Since the trace of (\ref{eq:f(R)eom}),
\begin{equation}
3 \square f' - \mathcal R\,f' - 2 f
 = T_\mathrm m\,,
\label{eq:f(R)eomtr}
\end{equation}
is an equation of motion for the scalar curvature $\mathcal R$ it is natural to promote it to the status of independent dynamical variable, the ``scalaron'': $\mathcal R=s$ \cite{Starobinsky:1980te}.
In so doing one converts (\ref{eq:f(R)eom}--\ref{eq:f(R)eomtr}) into a set of two second-order differential equations.

This scalaron can also be introduced right from the beginning by replacing the action (\ref{eq:f(R)act}) by the Dirac action
\begin{equation}
S_\mathrm S[g_{ij},s,\phi]
 = \frac{1}{2}
   \int\!\mathrm d^4x\,\sqrt{-g}\,
   [f(s) - \phi\,(s - \mathcal R)]
   + S_\mathrm m[\Psi,g_{ij}]\,,
\end{equation}
where $\phi$ is a Lagrange multiplier.
Now, the extremisation of $S_\mathrm S$ with respect to $s$ yields an algebraic constraint~: $\phi=f'(s)$\,, which can be harmlessly plugged back into $S_\mathrm S[g_{ij},s,\phi]$ yielding another action
\begin{equation}
S_\mathrm J[g_{ij},s]
 = \frac{1}{2}
   \int\!\mathrm d^4x\,\sqrt{-g}\,
   [f'(s)\,\mathcal R - (s\,f'(s)-f(s))]
   + S_\mathrm m[\Psi,g_{ij}]\,.
\label{eq:f(R)Jact}
\end{equation}
Extremising (\ref{eq:f(R)Jact}) with respect to $s$ and $g_{ij}$ yields the equations of motion (\ref{eq:f(R)eom}) (if $f''(s)\neq0$).
Note that the scalaron $s$ is not yet manifestly dynamical as its derivatives $\partial_i s$ do not appear in (\ref{eq:f(R)Jact}).
$S_\mathrm J[g_{ij},s]$ is the ``Jordan frame'' action of $f(\mathcal R)$ gravity ;
it falls into the broader category of scalar-tensor theories, see \cite{Damour:1992we}.
\bigskip

Eliminating the function $f'(s)$ in the term $\sqrt{-g} f'(s) \mathcal R$ in (\ref{eq:f(R)Jact}) by means of a conformal transformation will turn $s$ into an obvious dynamical variable \cite{Teyssandier:1983zz}.
Moreover it will lift the restriction $f''\neq0$\,, that is, it will render the Einstein limit well-defined.

Indeed, introduce the new metric
\begin{equation}
\tilde g_{ij}
 = f'(s)\,g_{ij}
\quad\Longrightarrow\quad
\mathcal R
 = f'(s)
   \left[
    \tilde{\mathcal R}
    + 3\,\tilde\square\ln f'
    - \frac{3}{2}\,(\tilde\partial\ln f')^2
   \right]
\label{eq:f(R)confR}
\end{equation}
(which imposes that $f'(s)$ be positive).
The action (\ref{eq:f(R)Jact}) becomes
\begin{equation}
\tilde S_\mathrm E[\tilde g_{ij},s]
 = \frac{1}{2}
   \int\!\mathrm d^4x\,\sqrt{-\tilde g}\,
   \left(
    \tilde{\mathcal R}
    - \frac{3}{2}\,(\tilde\partial\ln f')^2
    - \frac{s\,f' - f}{f'^2}
    + 3\,\tilde\square\ln f'
   \right)
   + S_\mathrm m[\Psi,g_{ij}={\tilde g_{ij}/f'}]\,.
\end{equation}
As announced, derivatives of $s$ now appear explicitly.
As for the term
\begin{equation}
\frac{3}{2}
\int\!\mathrm d^4x\,\sqrt{-\tilde g}\,\tilde\square\ln f'
 = \frac{3}{2}
   \int\!\mathrm d^4x\,\partial_i (\sqrt{-\tilde g}\,\tilde\partial^i\ln f')\,,
\end{equation}
it is a divergence which can be dropped.
Hence the final action is, after trading $s$ for a new field $\tilde\phi$~:
\begin{equation}
\tilde S_\mathrm E[\tilde g_{ij},\tilde\phi]
 = \int\!\mathrm d^4x\,\sqrt{-\tilde g}\,
   \left(
    \frac{1}{2}\,\tilde{\mathcal R}
    - \frac{1}{2}\,(\tilde\partial\tilde\phi)^2
    - V(\tilde\phi)
   \right)
   + S_\mathrm m[\Psi,g_{ij}=e^{-\sqrt\frac{2}{3}\tilde\phi}\tilde g_{ij}]\,,
\label{eq:f(R)Eact}
\end{equation}
where the potential $V$ and the new scalaron $\tilde\phi$ are given in terms of $s$ by~:
\begin{equation}
V(s)
 = \frac{s\,f'(s) - f(s)}{2\,f'(s)^2}\,,
\quad
\tilde\phi(s)
 = \sqrt\frac{3}{2}\,\ln f'(s)\,.
\label{eq:f(R)defphi}
\end{equation}
$\tilde S_\mathrm E[\tilde g_{ij},\tilde\phi]$ is the ``Einstein frame'' action, where Einstein's gravity is minimally coupled to the scalar field $\tilde\phi$ and non-minimally coupled to the matter fields $\Psi$\,.

The field equations obtained by extremising (up to boundary terms) $\tilde S_\mathrm E[\tilde g_{ij},\tilde\phi]$ with respect to $\tilde g_{ij}$ and $\tilde\phi$ reduce to~:
\begin{equation}
\tilde G_{ij} - T_{ij}
 = e^{-\sqrt\frac{2}{3}\tilde\phi}\,T^\mathrm m_{ij}\,,
\quad\hbox{where}\quad
T_{ij}
 = \partial_i\tilde\phi \partial_j\tilde\phi
   - g_{ij}
     \left(
      \frac{1}{2}\,(\tilde\partial\tilde\phi)^2
      + V(\tilde\phi)
     \right)
\end{equation}
and where, recall, $T^\mathrm m_{ij}=-\frac{2}{\sqrt{-g}} \frac{\delta S_\mathrm m}{\delta g^{ij}}$\,.
As they should, these equations are a rewriting of the Jordan frame equations of motion (\ref{eq:f(R)eom}) in terms of $\tilde g_{ij}=f'\,g_{ij}$ with $f'=e^{\sqrt\frac{2}{3}\tilde\phi}$\,.
Note that the equivalence holds if $f'>0$~;
note too that, as announced, the Einstein limit $f'\to1$ is well defined.

\section{\label{sec:mini}
Mini-superspace Hamiltonian formulation of $f(\mathcal R)$ gravity~: from the Ostrogradsky to the Einstein frame variables
}

We show here that the mini-superspace Hamiltonian formulation of $f(\mathcal R)$ gravity \`a la Ostrogradsky is (classically) equivalent to its formulation in the Einstein frame.
For better comparison with \cite{Ezawa:1998ax}, which claims the contrary, our formulation closely follows its authors'.

We restrict our attention to the sub-class of LFRW metrics of the type $ds^2=-N^2 dt^2+a^2 d\vec x^2$\,, where the lapse $N$ and the scalar factor $a$ are function of time $t$ only.
Hence the Lagrangian, $L[a,N]=\frac{1}{2}\,N\,a^3\,f(\mathcal R)$\,, is a function of
\begin{equation}
\mathcal R
 = 6 \left(\frac{\dot a}{N\,a}\right)^.
   + 12 \left(\frac{\dot a}{N\,a}\right)^2\,.
\end{equation}
As in \cite{Ezawa:1998ax} we introduce
\begin{equation}
Q
 = \frac{3\,\dot a}{N\,a}
\end{equation}
as an independent ``Ostrogradsky'' variable, so that the Dirac Lagrangian is
\begin{equation}
L_\mathrm O[a,Q,N,u]
 = \frac{1}{2}\,N\,a^3\,f(\mathcal R)
   + u \left(\dot a - \frac{N\,a\,Q}{3}\right)\,,
\quad\hbox{where}\quad
\mathcal R
 = \frac{2}{N}\,\dot Q + \frac{4}{3}\,Q^2
\end{equation}
and where $u$ is a Lagrange multiplier.
One checks that the vacuum Euler--Lagrange equations reduce to
\begin{equation}
Q\,\frac{\dot f'}{N}
- f'\,\frac{\dot Q}{N}
+ \frac{1}{2}\,f
- \frac{Q^2}{3}\,f'
 = 0
\quad\hbox{with}\quad
Q
 = \frac{3\,\dot a}{N\,a}\,,
\label{eq:minicons}
\end{equation}
which is nothing but the $(00)$ component of the field equations (\ref{eq:f(R)eom}).

The conjugate momenta of $a$ and $Q$ are
\begin{equation}
P
 \equiv
   \frac{\partial L_\mathrm O}{\partial\dot a}
 = u
\quad\hbox{and}\quad
\Pi
 \equiv
   \frac{\partial L_\mathrm O}{\partial\dot Q}
 = a^3\,f'(\mathcal R)
\quad\hbox{with}\quad
\mathcal R
 = \frac{2}{N}\,\dot Q + \frac{4}{3}\,Q^2\,.
\end{equation}
As in the toy model of Appendix~\ref{sec:toy} these relations cannot be inverted to give $\dot a$\,.
However the Hamiltonian $H^*_\mathrm O=P\,\dot a+\Pi\,\dot Q-L_\mathrm O$ is still well defined if we inject the constraint $u=P$ in $L_\mathrm O$\,.
Hence (cf.\ Eq.~(3.18) of \cite{Ezawa:1998ax}~;
see also \cite{Ezawa:2003wh})~:
\begin{equation}
H^*_\mathrm O
 = N
   \left(
    \frac{1}{3}\,a\,P\,Q
    - \frac{2}{3}\,\Pi\,Q^2
    + \frac{\Pi}{2}\,\mathcal R
    - \frac{1}{2}\,a^3\,f(\mathcal R)
   \right)\,,
\quad \hbox{where}\quad
f'(\mathcal R)
 = \frac{\Pi}{a^3}\,.
\label{eq:miniOHam}
\end{equation}
$\mathcal R$ is a known function of $\Pi/a^3$\,, once the function $f$ is given.
$H^*_\mathrm O$ is a function of $N$\,, $q_i=\{a,Q\}$ and $p_i=\{P,\Pi\}$\,.
Hamilton's equations
\begin{equation}
\frac{\partial H^*_\mathrm O}{\partial N}
 = 0\,,
\quad
\frac{\partial H^*_\mathrm O}{\partial p_i}
 = \dot q_i\,,
\quad
\frac{\partial H^*_\mathrm O}{\partial q_i}
 = -\dot p_i
\end{equation}
give back the Friedmann equation (\ref{eq:minicons}) and can be written as
\begin{equation}
H^*_\mathrm O
 = 0\,,
\quad
\{H^*_\mathrm O, q_i\}
 = \dot q_i\,,
\quad
\{H^*_\mathrm O, p_i\}
 = \dot p_i\,,
\end{equation}
where the Poisson bracket of two functions $A$ and $B$ of $(N,a,P,Q,\Pi)$\,, is defined as usual by
\begin{equation}
\{A,B\}
 \equiv
   \frac{\partial A}{\partial P} \frac{\partial B}{\partial a}
   - \frac{\partial A}{\partial a} \frac{\partial B}{\partial P}
   + \frac{\partial A}{\partial\Pi} \frac{\partial B}{\partial Q}
   - \frac{\partial A}{\partial Q} \frac{\partial B}{\partial\Pi}\,.
\label{eq:minidefPB}
\end{equation}
\\

In order now to transform the Ostrogradsky Hamiltonian (\ref{eq:miniOHam}) to an Einstein frame one we change the Ostrogradsky variables $\{(a,P),(Q,\Pi)\}$ into new ones, $\{(\tilde a,\tilde p), (\tilde\phi,\tilde\pi)\}$\,, such that, see (\ref{eq:f(R)confR}) and (\ref{eq:f(R)defphi})~:
\begin{equation}
\tilde a
 = \sqrt\frac{\Pi}{a}\,,
\quad
\tilde\phi
 = \sqrt\frac{3}{2}\,\ln\frac{\Pi}{a^3}\,.
\end{equation}
(This is the transformation proposed in (2.19b) and (2.21) of \cite{Ezawa:2009rh} but not the one suggested in the last section of \cite{Ezawa:1998ax}.)
In order to find the momenta $\tilde p$ and $\tilde\pi$ in terms of the Ostrogradsky variables, we impose the transformation to be canonical, that is, such that
\begin{equation}
\{\tilde\pi,\tilde\phi\}
 = 1\,,
\quad
\{\tilde\pi,a\}
 = 0\,,
\quad
\{\tilde p,a\}
 = 1\,,
\quad
\{\tilde p,\tilde\phi\}
 = 0\,,
\quad
\{\tilde a,\tilde\phi\}
 = 0\,,
\quad
\{\tilde\pi,\tilde p\}
 = 0\,,
\end{equation}
where the Poisson bracket is defined in (\ref{eq:minidefPB}).
The first series of Poisson brackets yield (a result at odds with Eq.\ (2.22) of \cite{Ezawa:2009rh})~:\footnote{where, clearly, the equation of motion $\frac{\partial H}{\partial p}=\dot q$ was incorrectly used.}
\begin{equation}
\tilde\pi
 = \frac{Q\,\Pi - a\,P}{\sqrt 6}\,,
\quad
\tilde p
 = -\sqrt\frac{a}{\Pi} (3\,Q\,\Pi-a\,P)\,.
\label{eq:minitrans}
\end{equation}
The second series of Poisson brackets is then satisfied.

In order now to express the Hamiltonian (\ref{eq:miniOHam}) in terms of the new variables we have to invert (\ref{eq:minitrans}).
This gives\footnote{See (\ref{eq:transO2E}) for the full-fledged version of this transformation.}
\begin{equation}
\left\{
\begin{aligned}
 &
a
 = \tilde a\,e^{-\tilde\phi/\sqrt6}\,,
\quad
P
 = -\left(
    3 \sqrt\frac{3}{2}\,\frac{\tilde\pi}{\tilde a}
    + \frac{\tilde p}{2}
   \right)\,
   e^{\tilde\phi/\sqrt6}\,, \\
 &
Q
 = -\frac{1}{\tilde a^2}
   \left(
    \sqrt\frac{3}{2}\,
    \frac{\tilde\pi}{\tilde a}
    + \frac{\tilde p}{2}
   \right)\,
   e^{\tilde\phi/\sqrt6}\,,
\quad
\Pi
 = \tilde a^3\,e^{-\tilde\phi/\sqrt6}\,.
\end{aligned}
\right.
\end{equation}
The Hamiltonian (\ref{eq:miniOHam}) therefore becomes
\begin{equation}
H_\mathrm E
 = \tilde N
   \left(
    -\frac{1}{12} \frac{\tilde p^2}{\tilde a}
    + \frac{\tilde\pi^2}{2\,\tilde a^3}
    + \tilde a^3\,V(\tilde\phi)
   \right)\,,
\label{eq:miniEHam}
\end{equation}
where $V(\tilde\phi)$ is given in parametric form by
\begin{equation}
V(\mathcal R)
 = \frac{1}{2\,f'^2}
   (\mathcal R\,f'-f(\mathcal R))\,,
\quad
\tilde\phi(\mathcal R)
 = \sqrt\frac{3}{2}\,\ln f'(\mathcal R)
\end{equation}
and where we have set $\tilde N=N\,e^{\tilde\phi/\sqrt6}$\,.
$H_\mathrm E$ is nothing but the Hamiltonian deduced from the Einstein frame action (\ref{eq:f(R)Eact}) (\ref{eq:f(R)defphi}) of $f(\mathcal R)$ gravity when reduced to minisuperspace~:
\begin{equation}
L_\mathrm E
 = -3\,\frac{\tilde a\,\dot{\tilde a}^2}{\tilde N}
   + \frac{\tilde a^2\,\dot{\tilde\phi}^2}{\tilde N\,\tilde a}
   - \tilde N\,\tilde a^3\,V\,.
\end{equation}
Hence, contrarily to the claim in \cite{Ezawa:1998ax} and \cite{Ezawa:2009rh} one can transform the Ostrogradsky Hamiltonian (\ref{eq:miniOHam}) into the Einstein one (\ref{eq:miniEHam}) by means of a canonical transformation.

\end{document}